# Internet of Tangible Things (IoTT): Challenges and Opportunities for Tangible Interaction with IoT

**Leonardo Angelini** [1,*], **Nadine Couture** [2], **Omar Abou Khaled** [1] **and Elena Mugellini** [1]

1. HumanTech Institute, University of Applied Sciences and Arts Western Switzerland; leonardo.angelini@hes-so.ch; omar.aboukhaled@hes-so.ch; elena.mugellini@hes-so.ch;
2. ESTIA LaBRI, UMR 5800; nadine.couture@u-bordeaux.fr
* Correspondence: leonardo.angelini@hes-so.ch; Tel.: +41-(0)26-429-6745

**Abstract:** In the Internet of Things era, an increasing number of household devices and everyday objects are able to send to and retrieve information from the Internet, offering innovative services to the user. However, most of these devices provide only smartphone or web interfaces to control the IoT object properties and functions. As a result, generally, the interaction is disconnected from the physical world, decreasing the user experience and increasing the risk of isolating the user in digital bubbles. We argue that tangible interaction can counteract this trend and this paper discusses the potential benefits and the still open challenges of tangible interaction applied to the Internet of Things. To underline this need, we introduce the term "Internet of Tangible Things". In the article, after an analysis of current open challenges for Human-Computer Interaction in IoT, we summarize current trends in tangible interaction and extrapolate eight tangible interaction properties that could be exploited for designing novel interactions with IoT objects. Through a systematic literature review of tangible interaction applied to IoT, we show what has been already explored in the systems that pioneered the field and the future explorations that still have to be conducted.

## 1. Introduction

Thanks to technological advances, electronic devices are spreading in our environments, seamlessly integrating in our everyday life. When in 1991, Mark Weiser, considered nowadays the father of ubiquitous computing, imagined that technology would disappear to our sight and integrate in our daily routines [1], probably he was not expecting that technology would have modified so profoundly our lives. Smartphones have brought ubiquitous computing to the mass, and indeed, nowadays, interaction with technology is becoming more frequent and intuitive. In parallel, in the context of Ubiquitous Computing, since 1999, researchers working in the domain of Internet of Things (IoT) are overcoming technological barriers and are investigating novel applications for connected objects in healthcare, transportation and smart-environments, which could bring many potential benefits for the human well-being [2]. Intel is forecasting for 2020 200 billions of IoT objects, i.e., 26 connected devices per person [3].

However, until now most IoT research focused on the technological challenges, often overlooking the important research issues of how humans should interact with this multitude of IoT objects [4]. Moreover, today typical interfaces for IoT objects are based on smartphones or web apps, which exploit few of the innate human skills that we would normally use to interact with physical objects. As a result, humans are often spending most of their time interacting with what Victor called a "Picture under Glass" [5], sometimes with severe social consequences [6]. Richer interactions that goes beyond the screen and that better exploit our bodily human skills can be imagined for our future [5,7,8]. In particular, richer interaction paradigms for IoT could also help the user in understanding and trusting connected objects, in order to exploit better IoT potentialities and benefits. To this purpose, we propose to study tangible interaction applied to the IoT, what we call "Internet of Tangible Things" (IoTT). The nature of the tangible interface for IoT objects could be twofold: either it can support immediate interaction in the periphery of the user's attention [9], freeing cognitive resources for the user's everyday activities, or it can support meaningful and unexpected interactions that stimulate reflection and the understanding of the system [10]. Possibly, these different interaction paradigms could exploit different levels of attention of the user, switching according to the context of the interaction.

In this context, few works have explored tangible interaction principles that can drive the design of IoT interfaces that are situated in the physical world. In this article, we propose a systematic literature review to analyze the principles that have been suggested or implemented in previous work and those that have been not yet explored. This article discusses also open challenges and opportunities for further research in this field.

Section 2 introduces the Context of this work, Section 3 presents related work in this field and, in particular, the state of the art of human-computer interaction with the Internet of things and the state of the art of tangible interaction. Section 4 presents the methodology of the systematic literature review. Section 5 presents the results of the review and Section 6 discusses the opportunities and open challenges emerged from the review. Section 6 concludes the paper and presents future work.

## 2. Context

Atzori et al. individuated four typical application domains for IoT: transport and logistics, healthcare, smart environments and personal and social applications [2]. Since this article focuses more on the interaction with IoT, where humans have an active role and IoT objects can offer useful services during daily routines, the typical application domains explored will be more related to smart environments and personal and social applications than to logistics and healthcare. Domestic environments, work environments and public environments are those that are mostly relevant for human-computer interaction with IoT. In these scenarios, not only objects can communicate through the Internet and their behavior can change according to network information, but the object behaviors can also be affected by other people interacting at distance. This is particularly evident for all the social applications where the Internet of Tangible Things can be used to strengthen social relationship ad distance [11,12]. As an example of an IoT system, Figure 1 presents a scenario were a user interacts with an IoT object, which is connected not only to a similar remote object through the Internet but also to other objects that can be located in proximity of the first object. It is worth noting that the behavior of an IoT object can be affected by other people interacting with connected objects, or by external services (e.g., weather services) that are difficult to identify and materialize for the user. Therefore, the typical scenario faced by a user of an IoT object is often very complex. The purpose of this study is therefore to understand if tangible interfaces are able to facilitate the understanding and control of these complex scenarios, increasing trustiness in IoT systems and supporting long-lasting interactions beyond technology obsolescence. As a starting point, we aim at looking at existing tangible properties identified and used in previous work to increase the user experience with IoT objects.

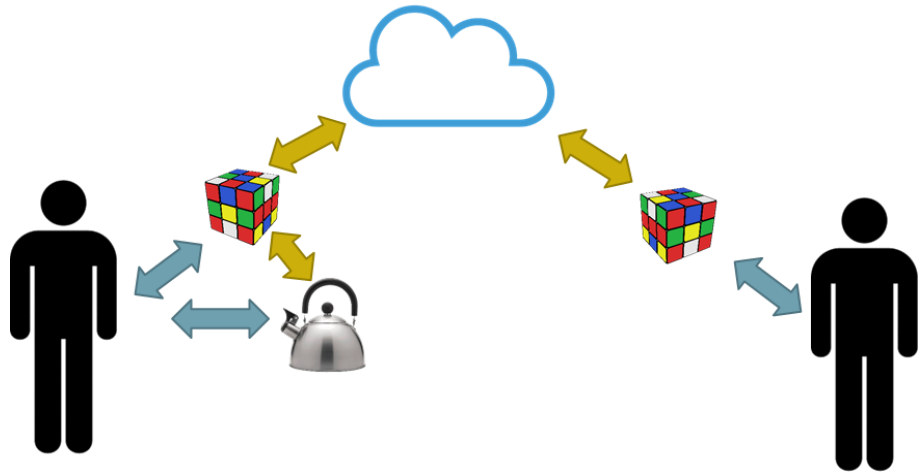

Figure 1. Interaction scenario for an IoTT system.

## 3. Related Work

This section presents related work in the two research fields that are closely related to the Internet of Tangible Things: the Internet of Things and Tangible Interaction. In particular, our initial analysis will focus on the current research on human-computer interaction applied to the internet of things (Section 3.1) and current trends of tangible interaction that seems promising for the internet of things (Section 3.2). We introduce then the new research field of the Internet of Tangible Things and discuss similar or related work (Section 3.3).

*3.1. Human-Computer Interaction with IoT*

The US National Intelligence Council defined the Internet of Things (IoT) as one of the six most "Disruptive Civil Technologies" [2]. The European Community is investing 192 million of euros in several different domains of IoT research through the Horizon 2020 programme for the years 2014-2017 [13]. Since the first definition of IoT, different technological platforms generated different related visions of what the IoT is. Atzori et al. identified three main visions: the first one is more "things-oriented" and respond to the initial trend of equipping everyday objects with RFID tags; the second vision, "Internet-oriented", relies on smarter objects, capable of fulfilling the full Internet Protocol (IP) stack; the third vision of IoT highlights the semantic reasoning that can be performed on IoT data [2]. The authors suggest that IoT stems from the intersection of these three visions. The Internet of Things is still a very actual research topic and many efforts are being devoted in order to standardize this broad ecosystem of interconnected objects: the Internet Architecture Board (IAB) has recently proposed a document for the definition of interconnected smart objects [14]. Similarly, in 2012, the ITU has proposed a first Recommendation for the definition of the Internet of Things [15].

A literature review of IoT shows that so far, most of the research on the Internet of Things focused on improving the hardware and software architecture that allow to IoT objects to better manage power consumption, optimize data collection and sharing as well as seamlessly communicate to each other, often without human intervention [16].

A search in the ACM Digital Library with the keywords (+("human-computer interaction" HCI) +("internet of things" iot)) produced 152 results, most of them concentrated in the last 3 years (from 2 results in 2010, to 14 results in 2013, up to 49 in 2016).

As pointed out also by Koreshoff et al.[4], the human is often neglected in IoT design and, only recently, some researches are conducted on the human-IoT interaction.

It is worth noting that interaction in smart environments and with the ubiquitous computer [17] has been already explored in depth. However, although these research fields share several challenges with the Internet of Things, the interaction with IoT introduces peculiar challenges, because of the worldwide interconnectivity between the different objects, people and virtual [18] and physical sensors, and because of their autonomous evolution according to exchanged information. In Designing Connected Products [19], Rowland et al. discuss the main challenges for improving the User Experience of consumers of IoT products. Rowland suggests that, because of the novelty of IoT in the consumer market, it is difficult to communicate to the user the conceptual model of how the device works and the related interaction model [19].

As a result, many IoT products fail to conquer a broad market, remaining niche products for geeks or experts because of functions that do not appeal to the mainstream user or configuration processes and interaction models that are not straightforward.

In order to design IoT objects that meet better users' needs and expectations, it is fundamental to analyze IoT also from a human-centered point of view. Koreshoff et al. readapted Atzori et al.'s IoT framework to address relevant research issues for Human-Computer Interaction (HCI) that are still open [4]. Koreshoff et al. conducted a literature review of HCI research and a review of IoT products [16]. The authors denote a lack of explicit involvement in HCI for IoT: most academic research focuses on specific application issues or in object interconnectivity and out of the 89 papers reviewed from the CHI conference and the Personal and Ubiquitous Journal, only 5 dealt with user inputs. For the 93 IoT commercial products reviewed, the authors show that most user interfaces are based on smartphone or web apps, or on screens integrated in the object.

Other attempts to promote user-centered design of the interaction with IoT devices can be found in Soro et al.'s article [20], who also propose an adaptation of Atzori et al.'s framework, and in Fauquex et al.' article [21], who propose a user-centered methodology for creating "people-aware" IoT applications.

Rowland et al. points out that IoT interaction models often suffer from their underlying technological complexity [19]. A typical property of IoT objects is to be connected to the Internet and to have a behavior that can depend from a dynamic network of remote connected entities. Therefore, the interaction designer for IoT should be able to cope with connectivity issues (Is the object still able to work when disconnected from the network? Will it respond to user inputs?) as well as with interaction models that depend on the autonomous communication with other connected objects (Did the object state change because of the sensed local environment of because of the object interaction? Is this object sending data to other objects? Which objects? Which data? For which people?).

Moreover, IoT objects are able to sense the external environment, collect sensitive information about the user and share it through the Internet. Therefore, a reliable control on sensed and shared data is particularly important for the user, who can otherwise lose the trust in the product or service. Bellotti and Sellen addressed these privacy issues with a framework for designing feedback and control in ubiquitous computing environment [22]. In particular, the framework addresses four system behaviors concerning feedback and control: when and which information is captured, how the information is elaborated, who has access to the information and for which purposes. While Bellotti and Sellen's framework supposes that actors that can access the information were humans, in IoT we face to a more complex scenario where also other IoT objects can access, elaborate and further share this information.

*3.2. Tangible Interaction and Perspectives for IoT*

This Section presents an analysis of current research in tangible interaction and highlights from literature the properties that in our opinion can be applied to the Internet of Things. These properties are:

P1. Meaningful representations and controls of the single IoT object connectivity status and IoT object interconnections, as well as of information capture, elaboration and sharing.
P2. Rich interactions that exploit the natural human skills, in particular exploiting haptic and peripheral interactions with IoT objects that are situated in the physical world
P3. Persistent physical representations that could last in case of power or connectivity outrage, allowing the user to control the state of an IoT object even when no Internet connection is available
P4. Spatial interactions that support collaborative setups with multiple IoT objects

P5. Immediacy and intuitiveness of the interaction, facilitating the understanding and control of IoT objects with minimal learning time
P6. Interactions with IoT objects that are integrated in the daily routines, that free users' cognitive resources and do not disrupt attention
P7. Facilitated reflections on IoT object meaning and working principles, as well as support for associating and sharing memories
P8. Long-lasting interactions with IoT objects exploiting emotional durable designs, to cope with electronic waste due to technological obsolescence.

The eight properties will be used as a framework for the meta-analysis of the papers selected in the systematic review.

Charlier analyzes in Chapter 8 of Designing Connected Products how different outputs and inputs can be combined in connected products to design appropriate interfaces and interactions [19]. Among the different interaction paradigms that can be explored for IoT, Charlier cites Tangible User Interfaces (TUIs) suggesting as advantages the usefulness for learning applications and music and the need of several tokens that can be lost as main disadvantages [19]. A recent research on tangible interaction, published at CHI'2016, suggests that richer interactions, beyond token manipulation, can be designed for IoT [23].

Since Fitzmaurice's Graspable User Interfaces [24] and Ullmer and Ishii's TUIs [25], Tangible Interaction has evolved much, embracing new disciplines and interaction paradigms [26]. Hornecker and Buur [26] individuated several advantages of tangible interaction over traditional interfaces: facilitates collaboration thanks to a shared interaction space and multiple access points (P4), exploits the spatiality and the user's proprioception (P4) as well as the interaction through the full-body (P2), makes use of expressive representations that help cognition (P1). Hoven et al. individuate three main aspects that characterize tangible interaction: the interaction with the physical world, the exploitation of the human skills and digital computation [8]. Although most digital interactions involve to some extent these aspects, the purpose of interacting with everyday objects exploiting human skills that are often ignored by traditional user interfaces makes tangible interaction different from other interaction paradigms and particularly interesting for interacting with IoT objects (P2). Hoven et al. individuated particular qualities of tangible interactive systems: they generally offer direct, integrated and meaningful representation and control of the digital information [8].

Since digital information is represented by a physical representation, which can be also used as a user control to manipulate such information, a particular property of Tangible User Interfaces is the persistence of the information even in case of power outage (P3). This a property that is particularly interesting for IoT, since IoT objects can run out of power or can be disconnected from the Internet (cf. [19]). The directness of the interaction and the coupling between action and perception [27] supports immediate interactions that are generally easy to learn and understand (P5). At the same time, objects can embody meaningful concepts and can support mindful interactions that make the user step back and reflect (P7) [10].

Among the different tangible interactive systems, it is possible to distinguish between two main categories: those that support immediacy and those that support mindful interactions.

This distinction stems from the Heidegger's definition of Ready-to-Hand and Present-at-Hand [28], which has been resumed by Dourish to introduce the principles of embodied interaction [29]. Tanenbaum et al. [30] extend this notion and map the two Heideggerian concepts to the Bolter and Grusin's concepts of transparent immediacy and hypermediation [31]. Tanenbaum et al. [30] discuss the typical Heidegger's example of the hammer, a tool that is typically Ready-to-Hand, until it breaks down and it appears to the user attentions, becoming Present-at-Hand: the hammer can also attract our attention without necessarily break down, for example because a detail of the grip reminds the user of past memories. Therefore, the authors suggest that the objects, besides their functional role, can be categorized according to a semantic line because they can become, present-at-mind. Also Hornecker [32] argued that tangible interaction has a hybrid nature and can support two different approaches, one that exploits affordances and user's previous knowledge to provide intuitive and seamless interactions and another that breaks the immediacy typical of direct manipulation in order to support reflection and the understanding of the system.

Tangible Interaction is particularly interesting for supporting peripheral interactions. Bakker and Niemantsverdriet [33] suggest that to better integrate digital interactions in our daily routines, the designers of interactive systems should exploit the full interaction-attention continuum. Indeed, the system should support both focused interactions that exploits all the users' cognitive and motor resources and lighter interactions that happen in the users' peripheral attention, which are more suitable when they are occupied in other tasks. Indeed, an important challenge that interaction designers have to face in ubiquitous computing is the overwhelming quantity of digital information and request for user input that is continuously presented to the user. This often disrupts the attention and concentration of the user, especially at work. Peripheral interaction tries to deals with this problem by designing interactions that can be performed through the user's peripheral attention, freeing user's cognitive resources for other tasks (P6) [9]. Bakker et al. suggests that peripheral interaction is particular useful for all those situations where the interaction could be integrated into the daily routine, without necessarily being in the center of the attention [9]. The designer of peripheral interfaces should also be able to take into account the context of the interaction, as well as take into account the personal preferences of the user [9].

While direct manipulation in touchscreen has been proved to be particularly intuitive and easy to learn, it requires visual attention and lacks of haptic feedback; tangible manipulation, instead, can be often performed without visual attention, relying on proprioception and haptic feedback for evaluating the result of the interaction [5].

The opposite approach to design valuable tangible interactions for long-lasting objects is bringing them to the center of the user's attention. In this case, the interaction should engage the user, stimulating reflection or emotional response [33].

Schmitz suggests that objects with life-like, often unpredictable, behaviors can enhance the longevity of the interaction, especially if coupled with zoomorphic or anthropomorphic affordances (P8) [10]. As in [10], Chapman proposes a framework for emotionally durable electronic devices [34]. The framework suggests that durable devices should have their own consciousness and should not be fully understood by the user (at least in the product exploratory phase), should make the user develop both attachment and detachment to the product and should carry signs of ageing and stories associated to the product itself. The purpose of Chapman framework is coping with the increasing waste of domestic electronic products, which get often abandoned by the user after few usages (P8). The association of stories to physical objects is a point that is often explored in tangible interaction to create attachment to a product. Hoven and Eggen [35] proposed an extension of Ullmer and Ishii's [25] Tangible Interaction Framework to take into account personal objects that are associated to personal memories (P7). Indeed, personal objects facilitate the association with digital media thanks to the pre-existing mental models. Tanenbaum et al. explored the relationship between narrative and objects through the Reading Glove, a wearable device equipped with a RFID reader to support storytelling with RFID tagged objects [30].

The ultimate challenge would be to combine both interaction paradigms (peripheral interactions and long-lasting focused interactions) in the same interface, allowing the user to switch between the two paradigms according to the context of use.

Among the emerging technologies that could potentially have an important impact in the design of tangible interactions with IoT objects, Shape-Changing User Interfaces (UI) [36] is probably the most interesting. While traditional tangible interfaces represent the state with a physical form that can be modified only locally by the user's physical manipulations, in Shape-Changing UI the physical state can be autonomously modified by the system to inform the user about a change in the digital state of the system. Ishii et al. called these systems Radical Atoms [37], which evolve the previous Tangible Bits systems [38] with the notion of transformation of the physical matter. Indeed, while Ishii's Tangible Bits interaction model [38] already involved a third loop feedback with actuation by the computer, this actuation was expected to happen only as a displacement of existing discrete objects, for example as in the PICO project [39]. In Radical Atoms [37], Ishii et al. suggested that also the form of the objects, and not only their position, could be dynamically modified, envisioning a world in which malleable digital matter could be shaped both from the user, through gestures and direct manipulations, or by the system, as feedback for the user. Several examples and applications of physical actuation for tangible interfaces have been presented, ranging from physical displacement of discrete objects on 2D (PICO) [39] or 3D space (ZeroN) [40] to physical actuation and input in 2.5D displays (inFORM) [41], inflation of deformable objects, recording and playing part motions of articulated objects (Topobo) [42]. The different types of actuation give birth also to different theories and names (e.g., Actuated Interfaces, Kinetic Interaction, Organic User Interfaces, etc.) who has been federated by Rasmussen et al. in 2012 into the more generic definition of Shape-Changing UI [36], who identified the following types of shape changes: orientation, form, volume, texture, viscosity, spatiality, adding/subtracting and permeability. Interestingly, they identified different purposes for shape-changing UIs and in particular hedonic aims (typical of Present-at-Hand systems) and functional aims (typical of Ready-to-Hand systems). Moreover, they identified different interaction models with the user for Shape Changing UI: only as feedback, as feedback to user's implicit input (e.g., feedback for automatically tracked user activity), as feedback to user's direct input or as a response to a remote user input.

This latter case is particularly interesting in the context of the Internet of Tangible Things, where connected objects could change their shape because of information received from the Internet, or because of the input of a distant user.

The implication on the user's perception of these objects behaviors are still unexplored. Recently, Rasmussen et al. evidenced also a lack of exploration in Shape-Changing UI that are not directly controlled by the system or by the user, but that fit in between this two extremes, negotiating the control with the user, or making the user control the state through implicit interactions [43]. It is worth noting that many IoT objects are designed to sense the user's activity and environment and to automatically act according to the detected activities.

Another final interesting trend in tangible interaction is the physical visualization of digital information, also called physicalization. Willet et al. highlights two techniques for embedded data representation [44]: they distinguish between embedded visualization, i.e., data represented in relation to the physical referent through augmented reality techniques, and embedded physicalizations, which have the purpose to present information that is directly integrated in the environment, in the physical referent or close to it. Shape-changing UI could be used also for this purpose. Jansen and Dragicevic [45] proposed a framework to support this kind of physical visualizations, discussing the challenges for representing digital data into 3D physical representation. Although their framework could be relevant also for IoTT, especially when large amount of data should be represented, our project focuses on a small amount of key IoT parameters that should be individually represented in a physical form to offer an immediate understanding and control over these parameters.

This analysis of the tangible interaction literature according to its two opposite and complementary approaches, which we identify in this article with the Heideggerian terms of Present-at-Hand and Ready-to-Hand, and of its emerging trends, suggests that tangible interaction offers many interesting possibilities for the design of new Human-IoT interfaces. The tangible interaction properties described in this section and summarized in P1) to P8) are promising for providing an easier understanding of the IoT objects and an increased trust in the IoT system. Through the systematic literature review, we aim at evaluating which of these properties have been explored until now and which properties still deserve further investigations.

*3.3. Internet of Tangible Things*

Tangible interaction in IoT is still mostly unexplored. Embedding the interaction in IoT objects is not a novelty and already in 2010, Kranz et al. [46] showed several examples in this field. The authors evidenced some challenges that are still actual nowadays, such as the risk of changing the nature of existing objects in order to add interactive capabilities, or vice-versa, the risk of hiding too much the interactive capabilities of the objects to the user eyes. Nevertheless, their analysis did not focus on the important implications of interacting with a network of objects that communicate to each other. The Web of Things, proposed by Guinard and Trifa [47] can be considered another close related domain. While in their first idea the Web of Things aimed at building web services for IoT objects, Mayer et al. [48] proposed a framework that associates semantic interaction primitives to different IoT object characteristics, with the purpose of supporting both web GUI and tangible interfaces, such as physical knobs and sliders. Although this was a valuable effort for bringing back IoT interfaces to the physical world, the richness of tangible interaction was still poorly exploited and the approach was mostly technology-oriented.

The novelty of the field of research explored in this article may be evidenced by the fact that "Internet of Tangible Things" is a new term introduced by Sarah Gallacher [49] only in January 2016 during the Second European Tangible Interaction Studio [50], with no previous reference in scientific literature. This term is adopted in this article to promote a shift towards the design of physical interactions with IoT. Indeed, the potential advantages and the related challenges are mostly unexplored until now in current research.

As a result of the systematic literature review presented in this article, it is worth highlighting some theoretical work that has been conducted recently in this field. Among the most relevant papers, it is worth citing the work of Ambe et al. [51], who analyzed three particular tangible IoT systems for connecting distant people and encourage social connection according to a selection of properties from Hornecker and Buur's [26] tangible interaction framework. In particular, they highlighted the importance of supporting user appropriation and personalization of IoT devices, in order to support long-term use of these devices. These properties are without doubt valuable for IoT objects that should foster social connection, but might not apply to other categories of IoT objects. For example, in VoxBox [52], a tangible questionnaire for collecting feedback during public events, appropriation was discouraged. Indeed, the purpose of the tangible questionnaire was to collect valuable feedback with intuitive but meaningful interactions related to users' event perception. In this case, while interactions should be playful, they should not be transformed just in a game. Instead, a tangible questionnaire should stimulate reflection on the proposed questions and facilitate discussions with other people.

The tangible properties explored or proposed and the approach for exploring them might vary consistently among different application domains, as well as research fields. In our literature review, we individuated different kinds of papers, from those having a user-centered approach, for example starting from understand the importance of personal objects for older adults [53], well before implementing the IoT systems, to more technology driven approach where the result not assessed with users[54]. The vision might also be slightly different between researchers with a computer science background and those with a product design background. For example, Knutsen et al. [55] through a review of commercial IoT products analyze the hybrid nature (tangible/intangible) of current IoT products, suggesting the term "Internet of Hybrid Products". Indeed, some of the commercial IoT objects reviewed presented a mix between a tangible interface and digital services based on smartphone or web apps. Although the idea is closely related to the IoTT concept, because of background the paper, was more focused on product and service design than on interaction design, discussing few of the aforementioned tangible interaction properties. For this reason, this paper was not included in the meta-analysis.

We found in literature interesting attempt to facilitate the interaction with IoT objects through tangible interfaces. Van der Vlist et al. [56] presented a system that allowed to create connections between IoT objects in the smart home through simple manipulations of tangible tokens that represent the IoT objects and of a central tile that represent the network hub. Domaszewicz et al. [57], presented a similar system where "an application pill" object allow to interconnect different IoT object and determine whether an IoT function can be obtained out of the IoT object that were in the range of the application pill. Ideally, different pills can be used to assess different functions. Those papers were considered as borderline since IoT objects were not augmented with tangible interfaces; instead, an additional tangible object was introduced to manipulate IoT object connections. Although this a valuable first attempt towards tangible interaction with IoT we did not include the papers in the analysis because, according to our eligibility criteria (cf. Section 4), the physical interaction was not embodied in the IoT object.

Besides all the work presented in Section 5, to the best of our knowledge, the only other relevant work that was not included in the literature review, is Physikit [26], a toolkit for the physical visualization of ambient data collected by an IoT sensor platform. Unfortunately, this article was not detected through our queries in the digital libraries (cf. Section 6.3).

**4. Methodology of the Systematic Literature Review**

The systematic literature review presented in this paper aimed at identifying previous work on tangible interaction and tangible user interfaces applied to the Internet of Things. The sources of the articles have been selected among the most popular digital libraries for Computer Science and in particular Human-Computer Interaction. The queries have been performed on three digital libraries: ACM Digital Library, IEEE Xplore and ScienceDirect.

A first research query aimed at individuating any article containing the keywords "tangible interaction" or "tangible user interface" and "internet of things" (Q1). Since several issues typical of the interaction with IoT were already addressed in previous work under the umbrella of "smart objects", we included in the search also articles containing the keywords "smart object" and "tangible interaction" or "tangible user interface" (Q2). Performing the queries on full text of the three aforementioned databases, we obtained 56 distinct results for Q1 and 29 distinct results for Q2. After removing duplicated papers obtained by different databases, we obtained 81 distinct papers.

The screening phase aimed at eliminating non-relevant papers based on objective criteria: book chapters without available full text (5), as well as keynote (1) and workshop proposal (6) papers, were discarded, in order to keep only articles with enough content. Moreover, papers containing "tangible interaction" or "tangible user interfaces" keywords only in citations () or as related work () were also discarded. Finally, to select only papers dealing with relevant content, the keywords "physical" or "tangible" and "connect", "network", "communication" or "relationship" were searched in the full text and analyzed in their context in order to discard all papers that were not dealing with physical objects and connected objects. After the screening phase, 34 papers were retained.

The eligibility phase aimed at individuating relevant papers that describe tangible interfaces for IoT systems or that discuss theoretical issues of tangible interaction with IoT. The eligibility criteria was defined as follow:

*The article talks about one or more physical objects, which are augmented with sensing, actuating, or information processing (storage, elaboration, communication) technologies and connected to other objects/people through a form of network technology (e.g., BT, WiFi, NFC, RFID). The user interacts with the object or system through a tangible interface, which exploits the physicality of the object.*

Two of the authors of this article have evaluated independently the eligibility for each paper. The two authors further discussed papers with discordant ratings. If no accordance could be found on the eligibility rating, a third author read the paper. At this point, eligibility was assigned according to a majority vote rule. After the first discussion 19 papers were declared as eligible, 12 as not eligible and 3 required the analysis of a third reviewer. After this phase, 18 papers were declared as eligible and 14 as not eligible. One paper [58] met the eligibility criteria, but an extended version [59] of the same article was also present in the pool; therefore, it has been excluded from the meta-analysis. This process has been conducted following a lightened version of the PRISMA review protocol [60]. The process is summarized in Figure 2.

18 papers have been analyzed to individuate which of the 8 tangible interaction properties presented in Section 3.2 have been explored already and which properties are still unexplored. The analysis allowed also individuating additional properties that were not listed among the aforementioned properties that we have previously individuated in tangible interaction literature, as well as open challenges that were highlighted in the papers. Each paper has been analyzed by two authors, who highlighted in the paper the phrases that were mentioning tangible properties. The author assigned each phrase to one of the eight properties or individuated a new property if this was not mentioned among the previously selected properties. Then they indicated for each article whether each of the 8 properties was present or not. Out of 144 attributions, only 19 cases (in average about one property per article) resulted with disagreement between the two reviewers and required further discussion. After discussion, reviewers reached agreement for all property attributions.

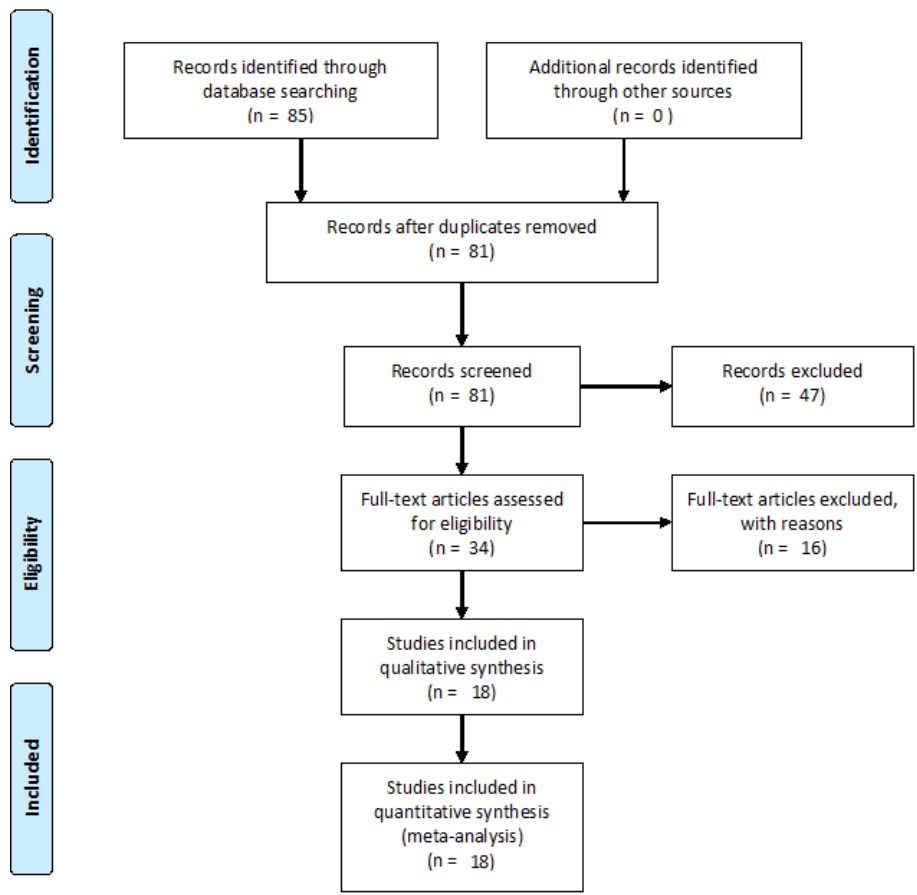

Figure 2. PRISMA 2009 flow diagram of the review process.

## 5. Results of the Meta-analysis

Since the purpose of the meta-analysis was mainly to individuate which tangible interaction properties have been suggested or implemented for the Internet of Things, this section presents how each property has been cited by previous work in this field. Additionally, we classified the articles found in the literature review according to different categories in order to understand whether these properties were merely suggested or they have also been implemented and tested with users.

Table 1 resumes the results of the meta-analysis for the 18 selected papers.

Table 1. Results of the meta-analysis.

| Paper | Type | P1 | P2 | P3 | P4 | P5 | P6 | P7 | P8 |
|---|---|---|---|---|---|---|---|---|---|
| TangiSense (Traffic) [59] | System, no user evaluation | X | | | X | X | | | |
| Active Forms [61] | Theoretical | X | X | | | X | X | | |
| AwareKit [62] | System, user evaluation | X | X | | X | X | X | | |
| CapNFC [54] | System, no user evaluation | X | X | | X | X | | | |
| Cognitive Objects [63] | Theoretical | X | | X | | X | X | | |
| Smart-home environment [64] | System, user evaluation | X | | | | X | X | | |
| TANGerINE [65] | System, no user evaluation | X | X | | X | X | | | |
| Expressing Intent [66] | System, no user evaluation | | X | | X | | X | X | X |

| VoxBox [52] | System, user evaluation | X | | | X | X | | X | |
|---|---|---|---|---|---|---|---|---|---|
| Invisible connections [53] | Theoretical, user interviews | X | | | | | X | X | X |
| IoT Owl [67] | System, no user evaluation | X | | | | | X | X | |
| Iyagi [68] | System, no user evaluation | | X | | | X | | | |
| TangiSense (Kitchen) [69] | System, user evaluation | | | | X | X | | | |
| Projected interfaces [70] | System, no user evaluation | X | X | | X | X | | | |
| RapIoT [71] | Toolkit | | | | X | | X | | |
| T4Tags [72] | Toolkit, end-user insights | | X | | X | X | X | X | |
| Technology Individuation [51] | Theoretical, analysis of user-evaluated systems | X | X | | X | X | X | X | X |
| Tiles [73] | Toolkit | X | X | | X | | X | X | |
| Total: 18 papers | 4 theoretical, 3 toolkits, 11 systems. 7 include user insights | 13 | 10 | 1 | 12 | 11 | 11 | 7 | 4 |

From the table, it is possible to notice that the most cited property was P1, with 13 occurrences out of 18 articles, while P3, i.e., exploring persistency of tangible interfaces, was the less mentioned (in only one paper, and marginally). Such sentences citing tangible interaction properties have been supported in different manners, in some cases they were just reported as possible benefits of tangible interaction, while in other cases they were supported by working implementations and in fewer cases by user evaluations. Indeed, we found 4 articles to be mostly theoretical (although one of them analyzed existing systems), 3 articles presented development toolkits and all the other articles (11) presented a system at different level of completion (at least 5 of these papers were published as work-in-progress). Insights from users were presented in 7 papers, but only 2 of the systems were properly tested with users [52,64]. Insights from users are presented also in two theoretical papers, i.e, Vaisutis et al. [53](with user interviews) and Ambe et al. [51] (with analysis of system tested with users), and in a toolkit paper [72](end-users tested a preliminary version of the toolkit, providing important insights).

Two theoretical articles, Active Forms [61] and Cognitive Objects [63], present concepts of particular IoT tangible objects. Active Forms are shape-changing physical objects that act as "gateways", i.e., user interfaces, for a Responsive Environment, i.e., a physical space permeated of ubiquitous, context-aware technology. Cognitive Objects, instead, aim at supporting humans and robots in their everyday routines, by means of specific functions embodied in their physical shape.

The other two theoretical articles focus on IoT objects for supporting social connection at distance. The first article [53] presents an interview to 6 older adults, conducted to investigate the possibility of augmenting everyday objects with IoT capabilities, in particular, for fostering social connection with distant relatives. This article also elicits human factors that should be considered in the design of such IoT objects. The second article [51] analyzes three existing IoT systems for fostering social connections according to some tangible interaction properties, with a focus on adoption, appropriation and habituation.

3 articles presented toolkits for developing IoTT objects. While Bellucci et al. [72] presented a toolkit for end-user programming of physical smart-tags (T4Tags) that can be attached to everyday objects; Mora et al. present two toolkits (RapIoT [71] and Tiles [73]) for facilitating the implementation of physical IoT objects (as well as the supporting cloud and embedded software) based on customizable hardware.

The other 11 papers present systems of various types: six papers present custom IoT objects (AwareKit, IoT Owl, Iyagi, Expressing Intent; Voxbox; TANGerINE); 3 papers present systems based on tabletops (2 systems based on TangiSense; TANGerINE); 6 use RFID or NFC tagged objects (2 systems based on TangiSense; CapNFC; AwareKit; Expressing Intent; Smart-home environment); one use a hybrid approach for augmenting IoT physical objects with light projection (Projected Interfaces).

More in detail, AwareKit [62] is a set of IoT objects connected to digital calendars that aims at facilitating meeting scheduling with colleagues. IoT Owl [67] is an IoT object with an owl shape that retrieves the frequentation of a canteen through another physical object (a stump) and suggests to office employees the right moment to go to eat. Iyagi [68] is a set of tangible objects to enhance bedtime routines for children through storytelling. Expressing Intent [66] presents a set of IoT objects animated by an own soul. The purpose of this article is discussing the importance of considering object animism, rich interactions and heterogeneous ecosystems while designing IoT objects. VoxBox [52] is a tangible questionnaire for collecting feedback during public events (e.g., attendants' appreciation). TANGerINE [65] is a system for supporting collaborative interactions on tabletops where users can transport information and make special actions through a custom-designed IoT tangible cube. Two other systems are based on the TangiSense platform, an RFID-enabled tabletop that augments tangible tagged objects with virtual agents. The two papers present two different application scenarios: for collaborative traffic management [59] and for suggesting recipes[69]. De la Guía et al. [64] presented a system designed to facilitate the management of smart homes through tangible NFC cards associated to different functions. Molyneaux and Gellersen presented Projected Interfaces [70], a system for augmenting smart tangible objects through environmental camera detection and light projection on the physical object. Grosse-Puppendahl et al. [54] presented a particular Capacitive NFC technology for augmenting everyday physical objects with IoT capabilities and enabling tangible interaction with these augmented objects.

In the following subsections, we will deepen the analysis of each property reporting the most relevant sentences from the 18 analyzed papers.

*5.1. P1: Meaningful Representations and Controls*

The first tangible interaction property proposed for the Internet of Tangible Things refers to the possibility of having meaningful representations and controls of the IoT object functions and/or parameters. The concepts of physical affordances and interaction embodiment are well known in the field of tangible interaction and, not surprisingly, this is the property with more occurrences among the selected papers. This concept was clearly highlighted by Scott-Harden for Active Forms [61]: "Active Forms are interactive products or devices that can render content thanks to perceptible changes to their physical form and appearance. The content is about the internal state of the products and about some of the application(s) and service(s) they support. The changes are a result of either user actions or internal actuators, both the physical form, such as shape or size, and the appearance such as colour or temperature can change."

Similarly, Cognitive Objects [63] "are embodied both in the task or action they are involved in and the environment. This is possible by holding semantical knowledge about themselves (such as the purpose of an object, how and when to use it, usage history, etc.)". Also Baraldi et al. with TANGerINE [65] are "interested in tangibles as the embodiment of some aspects of the interaction between the user and the domain of multimedia contents handled by the application."

A meaningful form for the object can be exploited to communicate to the user the IoT object main function. Hannula et al. [67] describes the IoT Owl as "a textile product that, via lighting and movement, displays the presence information from a sensor at a remote location, based on Bluetooth scan information.", specifying that "An owl was chosen as a character hence it is an animal, which is traditionally thought to be a symbol of intelligence. As a representation of a virtual friend it is so wise that it knows when people are present." For the Awarekit [62], Matviienko et al. highlight that "The unique shapes of the four modules are suggestive of the action or information they present.". Indeed, a bell was used to inform about reminders, a round clock for selecting hours, a rectangular block with five boxes for selecting the day of the week and a magnifying lens for the tooltip module. Meaningful representations and controls can also be embedded as part of more complex objects. For example in VoxBox [52] "The telephone handset that was used for the open question is also a very familiar interface that everyone knows what to do with, as was evidenced by the natural reactions to pick it up and say "hello" (even by very young children)." Voxbox "also incorporates a physical progress bar at the side. As questions are answered, a rubber ball is dropped one stage at a time through a transparent tube on the side of VoxBox."

Also in Projected Interfaces [70] it is highlighted that "The ability for objects to function as both input and output medium simultaneously enables scenarios such as objects that monitor their physical condition and visually display warnings if these are critical."

Such meaningful tangible representation can be useful to provide users of IoT objects with more intuitive controls on privacy settings. Indeed, Ambe et al. [51], while describing the Messaging Kettle, a system for supporting social interactions at distance through a device that detects and notify when someone is preparing tea with the distant kettle, underlines that "The Kettle Mate [an IoT separate object for detecting the kettle activity] can also simply be turned away, if there is a desire that its eye (temperature sensor) will not see the kettle boiling". At the same time, they add that "Users of the Messaging Kettle found the subtle glowing light indicating the other kettle is boiling to be an unobtrusive and yet enchanting representation of the boiling pot on the other end.", showing that the physical representation is not only meaningful for its intrinsic semantic value but has also a hedonic added value.

From the perspective of Mora et al. [65], whose objective is providing interaction primitives that are automatically recognized by their IoT development toolkit, "primitives become meaningful when they relate to the affordances of a specific object. Tiles can be used to make non-experts familiar with IoT concepts."

As an alternative to complex smart objects, several everyday physical objects or artifacts can also be augmented with tags in order to give physical forms to different functions or behaviors.

Grosse-Puppendahl et al. [54] augmented several objects with special capacitive NFC-tags to give them different functions: "approaching the rubber in distances up to 10 cm turns the lamp either on and off. Moreover, objects like a lighter or a whale are able to trigger specific lighting profiles like fire or ocean animations".

Limiting the physical embodiment physical cards with icons, de la Guía et al. [64] proposed task actions embodied in physical cards. "A tag (or more) is integrated inside the object or card depending on the size of the object; each tag describes a unique identifier. When the NFC reader in the mobile device is brought closer to the chosen representative object, the NFC tag inside is recognized by the NFC reader, and then the task is executed."

Vaisutis et al. [53] give a different perspectives on the meaning that can be associated with tangible objects: "Early work in tangible interaction emphasized the possibility of coupling digital and physical representations, while later work recognized the interweaving of the material/physical and the social aspects of interaction from a design perspective", highlighting from the interviews that "Some objects clearly symbolized tradition, status or prestige."

Finally, Ambe et al. [51] suggest that often allowing users to personalize the meaning associated to physical representation is a particularly valuable design choice. For example, "The Whereabouts Clock [8], is a mantelpiece clock-like device that depicts each family members' current location (based on their smartphone GPS position). Each family member initially registers different locations in their smartphone to refer to home, work or school (the Clock's regions).". The authors describe how different users characterized their specific association to work and home according to their specific daily lifestyle.

*5.2. P2: Rich interactions and human skills*

This tangible interaction property aims at promoting the exploitation of natural human skills, in particular, exploiting haptic interactions and rich manipulations with IoT objects that are situated in the physical world. While most interactions with nowadays IoT objects are constrained behind a screen and can be operated with a tap or swipe of one finger, our body can be used for much more complex interactions. Designers should be aware and try to exploit all our sensorimotor skills, both for perception (for example, using peripheral vision or haptics) and execution (for example, with full-body or bimanual interactions).

Also Brenton et al. [51] stress that "While objects are often manipulated with the hand, they are experienced by the entire body. Embodied interaction centres on the role of the body and bodily interaction in human understanding."

In their work Expressing Intent, Ng et al. [66] deserve a strong attention to rich interactions: "we felt that an overly digital interface doesn't leverage people's innate ability to interpret the expressiveness of their physical world. As a result, when we prototyped our objects we utilized the framework of Rich Interaction (Frens, 2006) and tried integrate form, interaction, and function into each product. Specifically, we have chosen to explore dimensions such as audio output, distance between objects by extending objects' natural affordances with digital abilities, thus creating opportunities for more expressive interfaces."

Also the authors of Iyagi [68] stress the importance of rich interactions for their interactive system for bedtime routines. As an example, they cite the Exploratorium of San Francisco: "Interactive spaces allow for rich interactions. This approach can be seen in spaces like the Exploratorium in San Francisco, CA where children explore complex concepts physically."

Scott-Harden suggests that an example of possible implementation of Active Forms [61] would use most of users' senses: "A household product that is designed to alert the user to an action or an event through interaction of the senses (sight, hearing, smell, taste, touch, balance and acceleration, temperature, kinesthetic sense, pain and also other internal senses)."

Several IoTT object implementations provide concrete examples of rich interactions. Matviienko et al. [62] highlight rich interactions for both input and output: "This networked toolkit provides availability information (day and week) for a single or a group of colleague(s) via different feedback modalities, such as light and sound, without need for a smart phone or desktop application. The modules in AwareKit, can be rotated, touched and connected [...]". Grosse-Puppendahl et al. [54], concerning the design of objects for blind users, say that "The first object we realized is a plastic card with braille text", adding that: "As the behavior of each object is very specific, we implemented context-sensitive audio hints for each object by shaking it". Also in TANGerINE[65] rich outputs are provided: "The result is stored, sent and translated in visual feedback by activating the LEDs as explained in our previous work. We implemented also a vibra-motor based layer to provide tactile feedback to the user." Finally, Molyneaux and Gellersen [70] elicit typical rich interactions that users can perform with their projected interfaces: "Direct manipulation of object (e.g. shaking ); Manipulation of object morphology; Manipulation of physical interaction components".

Also toolkits aims at supporting the definition of rich user interactions. Bellucci et al. [72] say that "custom events can be defined through tangible interaction with a programming-by-demonstration approach, in which the system learns new triggers by "watching" the user perform or demonstrate them (motion gestures, for example). Predefined gestures are provided, such as "shake," "swipe," and "up-down"." Similarly, Mora et al. [73] "[…] developed a set of input/output interaction primitives representing some of the most common manipulations performed on everyday objects that can be sensed with technology." In a table they list as input primitives touch, shake, and rotate, and as output primitives LED feedback and haptic feedback.

*5.3. P3: Persistency*

This property stresses that persistent physical representations could last in case of power or connectivity outrage, allowing the user to control the state of an IoT object even when no Internet connection is available. This would be a nice feature in comparison to smartphone interfaces that require connectivity to control the connected objects. Surprisingly, we could not find any specific reference to this property in the selected articles, although some systems (for example those based on tabletops) might inherently support this property. The closest argument to this property was provided by Möller et al. [63] while discussing the physicality property of Cognitive Objects (CO): "CO are no agents or digital representations, but physical objects incorporating affordances for humans and robots." suggesting that the interaction with cognitive objects could last beyond their numerical representation. Unfortunately, no practical implementation of Cognitive Objects has been presented to support this property.

*5.4. P4: Spatial interaction and collaboration*

Spatial interactions are typical of tangible interaction, since they exploit the physical environment to support not only collaborative setups with multiple IoT objects, but also spatial reasoning.

Spatial arrangement of different IoT objects are highlighted in AwareKit [62]: "The two components that facilitate interaction between the four aforementioned modules include tangible figures that can be assigned to particular colleagues and magnetic connectors that enable communication between the different modules. A user can activate a module by touching the tangible figure to the top of the module. The Day and Week modules can also be interfaced together using the magnetic connectors at one of four equidistant locations located at the bottom of the modules". Grosse-Puppendahl et al. [54] describe similar spatial interactions: "When the bottle is leaned in the way someone would do to pour out water [in the proximity of the lamp], the lamp starts filling up in the color of the bottle's contents." Similar arrangements are suggested also for Projected Interfaces [70], where the user can interact "with other smart objects in the environment (e.g. bringing another object closer)".

The support for collaboration is often stressed. In TANGerINE [65], authors highlight that "One of the main concerns when designing for collaboration is that of supporting distributed participation: in a synchronous setting such as a face-to-face collaborative story-making activity, a system should enable multiple users' simultaneous interaction. SMCubes can support this by allowing each child to use his own cube to interact with the system at the same time as other children", adding that "The spatial configuration of sensitive areas can also encourage different levels of engagement: a child in the Nearby Area can be involved in the activity peripherally, while a child in the Active Area is taking the leading role.". Tabletops are typical setups that support collaboration and Lebrun et al. [61] also stress that "tangible objects […] can be manipulated by one or several users".

The innate support for collaboration was evident also in VoxBox [52], a tangible questionnaire that was originally conceived for a single usage at time. Nevertheless, VoxBox allowed for "for one individual user or one group of users to interact at any time" and indeed among all the usages of the system, "61% were groups ranging in size from two to five people. In such group interactions, group members typically crowded around VoxBox, talked through their answers and worked together." The support for collaboration added several benefits: "none of the group members had to isolate themselves from their family or friends to give feedback, while the others waited. Instead it became a fun family or group activity that included everyone." Moreover "The observational study results showed that group input is a positive behavior and worked very well in informal event settings like the Fan Parks. It encouraged discussion and participation by all group members."

Ambe et al. [51] highlight more complex scenarios of spatial interactions that occur also across distance: "The Messaging Kettle augments an ordinary kettle with sensing and messaging capabilities to allow connection between geographically distant friends or loved ones while undertaking the simple routine of boiling water in the kettle. Two kettles in two distant homes are connected enabling each party to see when the other's kettle is on". In particular, they stress the importance of leaving the possibility to the user to customize the spatial arrangement of IoT objects: "The Messaging Kettle was then conceived to take advantage of spatial interaction strategies already in use in the elderly person's home".

Ng et al. [66] suggest that including several users in the interaction with IoT objects can enrich the overall experience :"By including multiple human actors in the mix, the objects form an interesting matrix for each actor's preferences, intent and interactions to be interpreted and harnessed in multimodal outputs"

Spatial interactions were supported also by all the toolkits: "RapIoT [71] supports applications that make use of several devices connected to the same gateway. This allow multiple users to interact with several devices placed in the same environment [...] Collaborative applications are then a concrete possibility: users can cooperate interacting with different devices for a common goal".

Bellucci et al. [72] stress that "Tags were envisioned not only as a way to open content and convey information, but also to enable interactions with other smart appliances." And for TILES [73], "Square modules can be used as physical pixels; they can be scattered in space or linked side by side for increasing information granularity."

*5.5. P5: Immediacy and intuitiveness*

Another property of tangible interaction that is often cited in literature is the ability to reduce learning time thanks to the immediacy and intuitiveness of the interaction. Thanks to the affordances provided by physical objects, tangible interaction is often considered to be "natural".

Designing natural and intuitive interfaces is one of the goals of Active Forms[61]: "Inspiration for the form has to come from our surroundings, notably nature; this is to ensure we end up with an intuitive device, where physical properties suggest usage." Cognitive Objects [63] have a similar purpose: "They reduce necessary a-priori knowledge to use and interact with an object, as this knowledge is distributed and directly provided by the respective object."

Intuitiveness of the interaction is particularly important for older adults. For the Easy Smart-Home System [64], authors note that "The interface is the means by which the user interacts with the system, therefore it is very important that they are easy to understand, simple and intuitive. One of our aims was to design user interfaces and interaction techniques which are easy to use for everyone." The result of the experience conducted with older adults confirmed that their goal was attained: "Users did not have any problems with the system; they intuitively performed their tasks without focusing on the system", "Interaction with the system is simple and intuitive" and "They do not need prior knowledge of the system or device to use it."

Similar positive results were shared by Matviienko et al.[62]: "Overall, participants found AwareKit [...] easy to use. The shapes of the modules were functionally intuitive. For example: "Tipping was easy and intuitive to do."", and by Baraldi et al. [65]: "our initial experience with users highlighted that the approach is very lightweight: especially on collaborative systems we found users to promptly understand the paradigm and use the cube in a natural way after some trial attempts."

Grosse-Puppendahl et al. [54] [54] described the following benefits for natural interaction: "A user can directly identify and combine the objects required for interaction. [...] the knowledge of object manipulations and movements can be exploited [...]"

Similar benefits are announced for VoxBox [52] "using tangible inputs with clear affordances, help people to immediately know how to use the device and reduce worries of not using it correctly", for Projected Interfaces [70] "our work on Projected Interfaces […] enable serendipitous bi-directional user interaction with smart tangible objects", and were highlighted often by Ambe et al. [51]: "Technologies that make use of existing household objects can utilise natural interactions in the home with those objects, without the need to explicitly indicate activity in some other way", "Utilising manipulation of existing objects for communication, and by making it simple to also not communicate, for example, by turning an object away is an example of a lightweight, legible tangible approach, that might be adopted in designing for the IoT."

Finally, the exploitation of affordances of existing objects was considered by Bellucci et al. [72]as a means to facilitate learning "Augmenting existing objects with sensing capabilities leverages their affordances to benefit from user interaction with the physical world: manipulating visible artifacts facilitates situated learning and epistemic actions".

*5.6. P6: Peripheral interaction*

In an overwhelming world of digital technologies that might continuously compete for the attention of the user, there is an increasing need of interactions with IoT objects that are integrated in the daily routines, that free users' cognitive resources and do not disrupt attention. For this reason, peripheral interaction [9] is gaining considerable attention in the HCI community and it deserves attention also for designing interactions for the Internet of Things.

Concerns about reducing the users' cognitive load were raised by Scott-Harden [61] for Active Forms: "There is a balance to be had between the cognitive load on the user, the selection of modalities, the media bandwidth and the user attention handling." He also added that "The integration of the Active Form in its environment is very important, in line with the work on calm and ambient technology".

The goal of Cognitive Objects [63] is supporting the user in daily routines. As such, the interaction with Cognitive Objects is integrated inherently in these routines: "Cognitive Objects support humans and robotic systems in the task execution. CO collaborate with humans, robots and the environment - proactively and situated."

Ng et al. [66], while proposing objects that behaves autonomously and that can interact with the user according to their personality, raise an important concern about the possible disruption of attention: "How this inflects their interaction and negotiation with them is also a rich area worth pursuing, especially when multiple objects start vying for attention and interaction from human actors. Will the eventual information overload discourage human actors from having these animistic objects inhabit their physical space, or does the organic interaction between objects remain attractive to users?". They further suggest other unanswered questions, which deserve further discussion: "One such issue would be the hierarchy of intents expressed - should all objects be able to interact and interrupt human actors at the same level, or are there those which might only be limited to a periphery? If so, which ones and why?"

One possible answer can be found in the words of Matviienko et al.[62], who suggest that "Tangible interactive systems are known to facilitate smooth transition of attention between background and foreground". Indeed, with AwareKit they explore the "feasibility of light as a medium to represent information about upcoming events in an unobtrusive manner", leveraging "ambient feedback mechanisms for usage in an office environment."

The integration of IoT objects in daily routines is not only important in office environments, but also in domestic contexts. Vaisutis et al. [53] analyzed the role of everyday objects in older adults' routines to understand which could be the role of IoT objects for fostering social connections: "They use objects in their routines and through their use, however mundane, they are wrapped into life-stories and become intertwined and connected in the intricacy of their owners lives.". In particular, the authors identified a category of "Objects of routine and comfort: [...] Some participants identified objects that supported their daily routines such as at mealtimes and making tea.". They also address a paradigm shift in tangible interaction research: "research has largely been based on discrete tangible and embodied systems and has not examined multiple networked objects embedded in everyday lives. As a precursor to this study the "habituated objects" that support the habits and routines of an elderly person were investigated".

Ambe et al. [51] stressed further these concepts, which are fundamental for supporting technology individuation: "In personal communication through IoT objects such as a messaging kettle, it is important that people can use and arrange those objects to fit into their own places and routines". At the same time, they also argue that new routines can be created around these new objects: "This has the effect of making and extending routines to new spaces, where they would otherwise not exist or exist in a different form. Every new piece of technology needs to have its place made in the home, within the many other existing objects, tools and routines." The same authors show that feedback integrated in everyday objects (which could be noticed through peripheral vision) may help establishing these routines: "in the case of the Whereabouts Clock, the designers could have easily implemented the clock interface as a mobile application but they opted instead to give the clock a tangible form, exploring what routines and behaviour would emerge by making information available in a situated manner at a particular place." Similar interactions were discussed for the Messaging Kettle: "The Kettle Mate exhibits a gently dynamic lava lamp like glow to signify that the kettle at the other person's home is currently boiling."

Krishnaswamy et al. [68] also tried to integrate their IoT system in domestic daily routines: "While considering the Iyagi system, it was important to consider how parents and children might move through their normal bedtime routine, largely uninterrupted. Voice interaction, and incorporation of normal bedtime objects would allow Iyagi to be integrated into a range of bedtime routines, while immersive stories and engaging activities would support parents in facilitating successful bedtime routines."

The three development toolkits also tried to support interactions integrated in everyday routines. Talking about IoT, the creators of RapIoT [71] argue that "By enabling seamless interconnection of people, computers, everyday objects and environments it promotes collaboration off the screen, into our everyday routines.". To this purpose, one of the device they created with the toolkit "provides visual warnings using a green and red LEDs to display whenever the air quality captured by the device is over or above the average value provided by other users."

Mora et al.'s TILES[73] are also designed for similar purposes: "Tiles Square are electronic modules, small enough to be easily attached or embedded in everyday objects can be used as ambient interfaces for non-intrusive information awareness."

Finally Bellucci et al. [72] discuss interesting insights collected while users tested the T4Tags platform: "The family articulated that the tag technology would be useful only if it could be easily integrated with the ecology of physical artifacts they already used to store, manage, and transmit information, such as calendars, to-do lists, notebooks, or recipes."

*5.7. P7: Reflection and memories*

This property highlights a different nature of tangible interfaces. Indeed, some tangible interfaces can stimulate user reflections, eventually promoting behavior change. In other cases, tangible interfaces can facilitate recollecting memories associated to objects. In both cases, the interaction is mindful and at the center of the user attention.

Sometimes this property emerge even in systems that are developed to be intuitive and easy to use. For example, Gallacher et al. [52] note that "Although VoxBox was designed to be playful and deployed in a fun vibrant setting, our findings suggest that the majority of respondents took the survey seriously and answered the questions with some reflection. People physically moved the sliders and spinners up and down before finally settling on a level they were happy with. They discussed the questions with others in their group before selecting their answer and even individual users were seen talking aloud to themselves about the current question and what their answer should be." Users were also aware that the tangible questionnaire was supporting their reflections: "After interacting with VoxBox many respondents reflected on their experience in a very positive way. For example, one person stated, "It's something really fun but it is useful and collects data too. It doesn't take too long and it's like a game. If you came up with a <normal> questionnaire I'd run away!""". Moreover, Gallacher et al. [52] add that "For many users VoxBox also sparked feelings of nostalgia and fond memories of tangible toys and objects from childhood", showing how simple tangible designs might easily support memory recollection, especially in the current trends of a world that is more and more digitalized.

Vaisutis et al. [53] stress that designing IoT for supporting social connections should take into account "How memories associated with objects might be enhanced with technology". Indeed, from older adults' interviews, they noticed that some objects "harbor significant memories". For example, "The stereo was not simply a device on which to play music but a reminder of the speech he gave at his daughter's wedding.". They further add that "In some instances the story attached to the object is what adds to its significance and this might easily be lost if the inheritor does not value or share the story."

Ambe et al. [51]"stressed that some objects can obtain special values, which can be associated to long-term reflections and use: "These arrangements over time will take in the novel technology and its uses, eventually endowing it with unique and personal value. This process shapes the technology to reflect one's self-identity and, symmetrically, shapes one's self-identity in response to the technology used.". While describing the user interactions with an emergency care pendant, they note that "For example, an older adult accepting mortality placed the pendant in the crucifix on her bedside table. Peculiar, it may seem, but this reminded her of her mortality at the same time as not having the idea consume her. Reflecting her faith and belief in a higher being, it gave her comfort."

And also "Another older adult placed a clothes peg near the emergency system terminal to remind himself to push the button daily to alert the telecare system that he is well. By helping his memory, he is also asserting that he is well and able to retain his independence, enabling him to stay in his own home and avoid being institutionalised." Also for the Messaging Kettle, the authors suggest that "the augmented kettle possessed straightforward and simple aesthetics that evoked memories and gave special meaning to its users."

Interactions that facilitate memory recollection were supported by the T4Tags of Bellucci et al. [72]: "In other cases, tokens afforded more long-lasting scenarios—they were linked to media content to augment physical objects, kept as memories, given as a gift, or used as a password."

Hannula et al. [67] designed the interactions of their IoT owl to support user reflection and behavior change: "For example, the IoT Owl can request the user to stand up and interact with it directly. The prototype forces the user to rise up by flapping its wings and changing the color scheme of the LED lights to red. The user can end the IoT Owls performance by tickling its claws. By forcing the user to move, the IoT Owl can help the user to take care of wellbeing and prevent the negative health effects of sitting too long." Mora et al. [73] also suggest that their platform can be used to build objects that support behavior change: "Square modules can be used to promote behavioral change. For example, a square module next to a water faucet can display over-average water consumption data in a household."

*5.8. P8: Long-lasting interactions and emotional bonding*

This last property highlights the need of long-lasting interactions with IoTT objects. Designing sustained interactions has not only the aim of increasing the user experience, but has also an ecological value for avoiding electronic waste related to IoT object obsolescence. While integrating objects within everyday routines could help to this purpose, sometimes a stronger emotional relationship with the object can be supported through tangible interactions. This can be achieved in different manners, for example giving to the object a personality, an own will or associating to an object a personal value, for example the relationship with a distant relative.

Vaisutis et al. [53] note that in general "People have strong emotional attachments to their things." From the interview with older adults, they could evidence in one occasion that "The emotional connection he has with the object extends much further than the object itself." While "On another occasion it was the words that weren't spoken that indicated the depth of emotion associated with the objects." They also reported that "Some participants freely expressed joy and enthusiasm as they explained what their chosen objects meant to them." The authors end up with recommendations for supporting the design of IoT objects that foster social connection: "design of such technology needs to be framed by an understanding of: 1) The underlying emotions that are attached to and the social relations that are facilitated by objects; 2) The kind of communications desired (if any and with whom) and how this might be best facilitated by objects; […]".

The purpose of Ambe et al.'s article [51] was directly linked to this property: "In exploring how tangible interaction with augmented everyday objects unfolds in a home setting, building on the emotional value that people sometimes place on cherished objects and routines, we aim to contribute to the area of tangible and embodied interaction and the design of the Internet of Things."

From the analysis of the three previous systems, they could individuate several elements that allow developing emotional attachment with IoT objects: "Technology individuation makes the technology distinctly personal as it embodies personal history, emotional attachment, and time and effort spent in personalisation that reflects the owner's identity and relations". They also add that "The more an object is personalised and used to express the self, the more the emotional bond with that object strengthens". Specifically, talking about the Messeging Kettle, they state that "In both trials, we found the kettle is no longer just an object used for boiling water. It now represents the loved one from far away making tea" and that "The additional layer of interaction in the Messaging Kettle providing additional opportunities for communication, connection and emotional attachment, endowing the technology and its content with more meaning and special value. For Messaging Kettle and Whereabouts Clock users, the object is no longer just an everyday object but a connection to loved ones and a feeling of togetherness." The attachment can also be supported by daily routines: "one older user was quite happy when she replaced her old kettle with a new one, not being attached to the kettle itself but rather to the tea making routine. [...] attachment to a particular object may develop through using it over time and through memories invested in its use." Such routines can support sustained use and in the end, overcome object obsolescence: "The simple routine of making tea and each other's motivation to stay connected has sustained their use of the Messaging Kettle and strengthened the bond to the distant one. This enduring use characteristic of the Messaging Kettle reflects the call for durable, long lasting systems sustainable design".

Ng et al. [66] developed a whole point on the importance of giving IoT objects a personality, i.e., Object animism: "Increasingly, objects, machines and technology is infused with anthropomorphic elements that help enhance human interaction with them (Seehra, et al, 2015). Such designed animism, for Laurel, "forms the basis of a poetics for a new world." (Laurel, 2008: 252) whereby said objects' elicit divergent behaviors as an effect of interaction (Zuckerman and Hoffman, 2015). These additions are not merely functional, but incorporate the dimension of empathy and enchantment that makes human-computer interactions alluring."

Hannula et al. [67] tried to incorporate such principles in the design of the IoT Owl: "The IoT Owl was created to have a personality", "It is shaped like a character to make the users feel like it has a personality." Some particular interactions helped achieving this goal: "During the working process we created moving wings for the IoT Owl. Almost by accident the noise of the servo motors sounds like the IoT Owl is trying to communicate with the user verbally."

## 6. Discussion

### 6.1. Trends and Perspectives of the IoTT

As first discussion argument, we would like to deepen the analysis of articles found during the systematic literature review, in order to determine a general trend for the research field of the Internet of Tangible Things. While the buzzword of the Internet of Things has recently gained traction in research and in the industry, with no exception for the field of Human-Computer Interaction, some of preliminary works we found date back to 2008-2009. Older technological platforms relied often on tabletop setups (as support for tangible interaction) and on RFID and NFC (as network technology for the Internet of Things), which limited the communication capabilities and the intelligence embodied in IoT objects. Recent technological advancements and the diffusion of prototyping platforms, which make easier embedding technology in small-sized battery-powered everyday objects, is enhancing the capabilities of IoT objects. Big technology players, such as Intel, are also pushing this market and its technological advancement, supporting researchers as well as DIY and maker communities. As a result, not only objects gain sensing capabilities that were impossible with previous RFID and NFC IoT technology, but they also enable richer interactions with the user. Thanks to the diffusion of prototyping platforms, we noticed a flourishing of work-in progress papers for developing IoTT objects. We believe that designing objects that embody tangible interfaces for interacting with the object itself and with the larger amount of information provided by the Internet of Things is the aim of the IoTT field of research. The concepts of Active Forms [61] and Cognitive Objects [63] partially supported this research question. Unfortunately, it seems that no concrete effort has been pursued to implement these concepts. On the other side, we found three toolkits that aim at facilitating the development of IoTT objects, not only for engineers and designers, but also for end-users with no programming skills. Considering the large amount of work found in the last years in this field and the continuous traction of the larger field of the Internet of Thing, we expect an increasing number of researchers approaching the field of the Internet of Tangible Things.

Besides the technological aspects linked to the development of IoTT objects, we would like to encourage a user-centered approach for the design of the interaction with the IoTT. Indeed, few of the works found in the systematic literature review followed a user-centered approach. The work of Vaisutis et al. [53] and Ambe et al. [51] are exemplar in this direction. Starting respectively from user-interviews and from the analysis of long-term user evaluations of previous systems, they derived recommendation for the design of IoT objects that support social connection and that integrate in daily routines. The work of Bellucci et al. [72] shows the benefits of conducting exploratory evaluations with end-users and refining the system according to their feedback. Besides these articles, only two papers [52,64] provided a formal user evaluation and reported user insights, while most other work provided just preliminary evaluations or no evaluation at all. Since this is a new field of research, we can argue that most of the works are still in an exploratory or work-in-progress phase and further evaluations will be published in the future. However, proper user-evaluations should be conducted on a long-term basis and this is a challenging task (even in the context of funded projects) as highlighted by Bellucci et al. [72].

*6.2. Discussion of Tangible Properties*

As shown in Table 1, most properties were cited by several papers, with the only exception of the persistency property, which was almost not mentioned at all. It is worth noting that there was a general agreement among reviewers about the presence or non-presence of properties in each paper, and disagreement was always solved after discussion (cf. Section 4). Reasons of disagreement were related often to properties that were closely related. For example, P1, i.e., the use of meaningful representations and controls of IoT properties or functions, implies often P5, i.e., an interaction that is intuitive and easy to learn. Both properties rely on physical affordances and interaction embodiment, which are one of the pillars of tangible interaction. The little nuance that can be distinguished between P1 and P5 is that the meaningful representations and controls rely often on metaphors and on the semantics provided by the context of the application, while intuitiveness and immediacy can be achieved based on low-level concepts, such as naïve physics [74] or embodied metaphors [75]. Some authors referred to intuitive and immediate interactions as natural interactions. However, "natural" is an ambiguous term especially when referred to interaction, as Norman [76] underlined in a famous paper. Indeed, natural might also allude to the fact that the interaction is exploiting our innate natural human skills, such as our ability to manipulate physical objects. This was also a reason of initial disagreement with one reviewer, who considered natural interactions as related to P2 (rich interactions) instead that to P5. Rich interactions (P2), such as complex object manipulations, possibly performed with two-hands or with the whole-body, or such as interactions that exploits unconventional human senses, are often not immediate at all and in many cases require learning new skills. Users would often prefer immediate interactions such as swipes and taps [77,78], also because they are habituated to this kind of interactions with the omnipresent smartphone. Nevertheless, aren't the most engaging human activities, such as playing an instrument, practicing a sport, or driving a car, also those that require the longer time for learning? At the same time, those activities that require much of our attention during the learning phase, after some habituation, start to disappear into our daily routines and we free cognitive resources that can be used for other tasks (P6). In this context, peripheral interaction (P6) has the role to exploit these free resources in an obtrusive manner, allowing interacting with IoT objects while performing other activities. Obtaining such goal is not trivial; it requires a good knowledge of human brain sensorimotor capabilities and may imply exploiting unconventional interaction modalities (P2).

On the other side of the spectrum of tangible interaction's hybrid nature [32], we have interfaces that might not be so straightforward and that may require all our attention. These interfaces can stimulate reflection (P7), and even promote behavior change. Physical objects might also have personal meaning and can be linked to personal memories (P7). In this case, the meaningful representations (P1) are not tied to the intuitiveness (P5) anymore, but are rather associated to personal memories. And whenever the object is tied to personal memories (P7), there is often an emotional attachment with the object (P8). But how an emotional attachment to a new IoT object, with no memory associated, can be supported? How can we make the interaction with IoT objects long-lasting? Schmitz [10] suggests that animating objects with personality and providing unexpected behaviors can help creating an emotional attachment with the user. Long-lasting ephemeral interactions are also explored by the Expressing Intent [66] project, which animated IoT objects with different personalities. Nevertheless, the authors raised an important question. Should all objects be able to expose such personalities? Should they be able to prompt the user at their discretion, possibly disrupting her attention (P6)? The authors suggest that some objects can expose such behaviors while others should not, depending of their role in our life. However, Bakker and Niemantsverdried suggested that we should start exploring the full interaction-attention continuum [33], therefore, some objects should catch the user's attention in appropriate contexts, while supporting our daily tasks during normal routine.

Spatial Interaction (P4) is also a common property in the context of tangible interaction. This property can also assume some nuances. Indeed, it is well-known that most tangible interfaces, offering several distributed input can foster collaboration. At the same time, collaboration can facilitate discussion, which is tied implicitly to the ability to make the user reflect (P7). Similarly, interaction in space and, in particular, object manipulation in space facilitate spatial reasoning; therefore, spatial interaction can stimulate reflection even in non-collaborative setups. At the same time object manipulation in space requires performing rich interactions (P2). This close link between P2 and P4 was one of the sources of confusion among reviewers. Finally, spatial interaction get even more complicated with the IoT, especially when different users start collaborating and communicating at distance. Are the aforementioned properties still valid? Further studies might be required in this case.

Finally, it is worth discussing the Persistency property (P3), the only one for which we found almost no reference in related work. Although the possibility to interact with an IoT object even without Internet connectivity (or even when the object is not powered anymore) might seem uniquely a technological challenge, this property might have stronger implications on the user experience with IoT objects. Indeed, such property can increase the user trustiness on the IoT objects, allowing complete control even in case of temporal failure. In the end, it can potentially increase the perceived robustness of the system. Therefore, we encourage explorations of this property in future work.

*6.3. Limitations of this Study*

Although the purpose of a systematic literature review is including in the analysis all relevant work in a specific field of research, we are aware that some relevant works could be missing from the selected papers. In order to limit the number of papers that could be excluded from the analysis, we conducted the queries for the selected keywords on full-text rather than just on meta-data or on the abstracts. We also included a query for smart-object, since the IoT keyword could be missing in non-recent relevant work. Unfortunately, after completing the analysis, we realized that a relevant paper that we already cited from the generic literature review was missing. Indeed, this paper was not among the 81 papers retrieved through the queries, although it do include the internet of things and tangible interaction keywords. After double-checking the query, we concluded that the search on full-text might not work as expected on the ACM Digital Library. As a result, other relevant papers might be missing from our analysis. We could have included the Physikit paper in the analysis, in a second time, exploiting the "Other Sources" branch of the PRISMA flow diagram; however, this would not guarantee the completeness of the review. Therefore, we preferred to stick to systematic methodology and we did not analyze the paper.

The selection process and definition of tangible properties could be limited by an a-priori selection based on our knowledge of the field and on a non-systematic literature review. The properties could have been derived from the systematic analysis of the papers, but it was much easier to identify properties with an a-priori knowledge of what we were looking for than just identifying potential phrases that could be discussing potential common tangible properties. In the end, all but one tangible properties were clearly identified in previous work, showing that most of them were particularly relevant and already cited in the field. As discussed in Section 6.2, we also detected some interdependence of tangible properties and nuances between them, which caused in some cases an initial disagreement after individual property identification. These disagreements were solved after discussion between the authors (the respective reviewers of the papers). Nuances and interdependences between properties are still present, but we argue that this is an intrinsic characteristic of tangible interaction. Indeed, as highlighted by Hoven et al. [8], also tangible interaction has no clear boundary but that has many concepts and properties that should be considered to improve the interaction with digital information.

Finally, this paper does not discuss the possible associations between tangible interaction properties and IoT properties (e.g., the object connectivity state or the information capture ability) and how these associations can improve the user experience with IoTT objects. Such analysis, which is without doubt relevant, will be conducted as future work.

**5. Conclusions**

This paper presented a systematic literature review of tangible interaction applied to the Internet of Things. We defined this research field "Internet of Tangible Things" (IoTT). This work aims at lying the basis for future research in the field, which seems promising considering the increasing number of work published in the last years. We proposed 8 tangible interaction properties and we analyzed the existing work in the field according to these properties. Out of 81 distinct papers containing keywords related to tangible interaction and IoT, we selected 18 relevant papers, which included 4 theoretical papers, 3 papers presenting toolkits for developing IoTT objects and 11 papers presenting IoTT systems. We found that 7 of the 8 properties were well represented in the literature and we provided examples of how the properties were mentioned, either as theoretical citations, as implementation solutions, or as user insights. We believe that such work and, in particular, the tangible interaction properties that we individuated can be inspiring for the design of interaction with new IoT objects. To this purpose, we promote a shift towards the design of interactions that are embodied in everyday IoT objects and that offer richer experiences compared to current graphical user interfaces for IoT.

As future work, we would like to test if the tangible interaction properties that we elicited in this paper can be useful to design better interactions with IoT objects. Moreover, we would like to understand how tangible interaction properties can be exploited to improve the understanding and control of IoT properties. To this purpose, a design workshop [79] with researchers and practitioners in the field has been conducted during the Third European Tangible Interaction Studio. Results of this activity will be analyzed and discussed in a future paper.

**References**

1. Weiser, M. The computer for the 21st century. *Scientific american* **1991**, *265*, 94-104.
2. Atzori, L.; Iera, A.; Morabito, G. The internet of things: A survey. *Computer networks* **2010**, *54*, 2787-2805.
3. Intel iot gateway. http://www.intel.com/content/dam/www/public/us/en/documents/product-briefs/gateway-solutions-iot-brief.pdf (April 2016),
4. Koreshoff, T.L.; Leong, T.W.; Robertson, T. In *Approaching a human-centred internet of things*, Proceedings of the 25th Australian Computer-Human Interaction Conference: Augmentation, Application, Innovation, Collaboration, 2013; ACM: pp 363-366.
5. Victor, B. A brief rant on the future of interaction design. 2011.
6. Jägemar, M.; Dodig-Crnkovic, G. In *Cognitively sustainable ict with ubiquitous mobile services: Challenges and opportunities*, Proceedings of the 37th International Conference on Software Engineering-Volume 2, 2015; IEEE Press: pp 531-540.
7. Marti, P. The subtle body. *Eindhoven Technical University Library, ISBN* **2014**, 978-990.
8. van den Hoven, E.; van de Garde-Perik, E.; Offermans, S.; van Boerdonk, K.; Lenssen, K.-M.H. Moving tangible interaction systems to the next level. *Computer* **2013**, *46*, 0070-0076.

9.  Bakker, S.; van den Hoven, E.; Eggen, B. Peripheral interaction: Characteristics and considerations. *Personal and Ubiquitous Computing* **2015**, *19*, 239-254.
10. Schmitz, M. In *Concepts for life-like interactive objects*, Proceedings of the fifth international conference on Tangible, embedded, and embodied interaction, 2011; ACM: pp 157-164.
11. Kowalski, R.; Loehmann, S.; Hausen, D. In *Cubble: A multi-device hybrid approach supporting communication in long-distance relationships*, Proceedings of the 7th International Conference on Tangible, Embedded and Embodied Interaction, 2013; ACM: pp 201-204.
12. Brereton, M.; Soro, A.; Vaisutis, K.; Roe, P. In *The messaging kettle: Prototyping connection over a distance between adult children and older parents*, Proceedings of the 33rd Annual ACM Conference on Human Factors in Computing Systems, 2015; ACM: pp 713-716.
13. Commission, E. Research & innovation. Internet of things. Digital single market. https://ec.europa.eu/digital-single-market/en/research-innovation-iot
14. (IAB), I.A.B. *Rfc 7452, "architectural considerations in smart object networking"*; 2015.
15. ITU-T. Recommendation y.2060: Overview of the internet of things. **2012**.
16. Koreshoff, T.L.; Robertson, T.; Leong, T.W. In *Internet of things: A review of literature and products*, Proceedings of the 25th Australian Computer-Human Interaction Conference: Augmentation, Application, Innovation, Collaboration, 2013; ACM: pp 335-344.
17. Schmidt, A.; Kranz, M.; Holleis, P. In *Interacting with the ubiquitous computer: Towards embedding interaction*, Proceedings of the 2005 joint conference on Smart objects and ambient intelligence: innovative context-aware services: usages and technologies, 2005; ACM: pp 147-152.
18. Sakaki, T.; Okazaki, M.; Matsuo, Y. In *Earthquake shakes twitter users: Real-time event detection by social sensors*, Proceedings of the 19th international conference on World wide web, 2010; ACM: pp 851-860.
19. Rowland, C.; Goodman, E.; Charlier, M.; Light, A.; Lui, A. *Designing connected products: Ux for the consumer internet of things*. " O'Reilly Media, Inc.": 2015.
20. Soro, A.; Brereton, M.; Roe, P. The messaging kettle: It's iotea time. **2015**.
21. Fauquex, M.; Goyal, S.; Evequoz, F.; Bocchi, Y. In *Creating people-aware iot applications by combining design thinking and user-centered design methods*, Internet of Things (WF-IoT), 2015 IEEE 2nd World Forum on, 2015; IEEE: pp 57-62.
22. Bellotti, V.; Sellen, A. In *Design for privacy in ubiquitous computing environments*, Proceedings of the Third European Conference on Computer-Supported Cooperative Work 13–17 September 1993, Milan, Italy ECSCW'93, 1993; Springer: pp 77-92.
23. Steven Houben; Connie Golsteijn, S.G., Rose Johnson, Saskia Bakker, Nicolai Marquardt, Licia Capra , Yvonne Rogers. Physikit: Data engagement through physical ambient visualizations in the home. In *CHI'16*, 2016.
24. Fitzmaurice, G.W. Graspable user interfaces. Citeseer, 1996.
25. Ullmer, B.; Ishii, H. Emerging frameworks for tangible user interfaces. *IBM systems journal* **2000**, *39*, 915-931.
26. Hornecker, E.; Buur, J. In *Getting a grip on tangible interaction: A framework on physical space and social interaction*, Proceedings of the SIGCHI conference on Human Factors in computing systems, 2006; ACM: pp 437-446.
27. Wensveen, S.A.; Djajadiningrat, J.P.; Overbeeke, C. In *Interaction frogger: A design framework to couple action and function through feedback and feedforward*, Proceedings of the 5th conference on Designing interactive systems: processes, practices, methods, and techniques, 2004; ACM: pp 177-184.

28. Heidegger, M. *Being and time: A translation of sein und zeit*. SUNY Press: 1996.
29. Dourish, P. *Where the action is: The foundations of embodied interaction*. MIT press: 2004.
30. Tanenbaum, J.; Tanenbaum, K.; Antle, A. In *The reading glove: Designing interactions for object-based tangible storytelling*, Proceedings of the 1st Augmented Human International Conference, 2010; ACM: p 19.
31. Bolter, J.D.; Grusin, R. Immediacy, hypermediacy, and remediation. *Remediation: Understanding new media* **1999**, 22.
32. Hornecker, E. In *Beyond affordance: Tangibles' hybrid nature*, Proceedings of the Sixth International Conference on Tangible, Embedded and Embodied Interaction, 2012; ACM: pp 175-182.
33. Bakker, S.; Niemantsverdriet, K. The interaction-attention continuum: Considering various levels of human attention in interaction design. *International Journal of Design* **2016**, *10*.
34. Chapman, J. *Emotionally durable design: Objects, experiences and empathy*. Routledge: 2015.
35. Van Den Hoven, E.; Eggen, B. Tangible computing in everyday life: Extending current frameworks for tangible user interfaces with personal objects. In *Ambient intelligence*, Springer: 2004; pp 230-242.
36. Rasmussen, M.K.; Pedersen, E.W.; Petersen, M.G.; Hornbæk, K. In *Shape-changing interfaces: A review of the design space and open research questions*, Proceedings of the SIGCHI Conference on Human Factors in Computing Systems, 2012; ACM: pp 735-744.
37. Ishii, H.; Lakatos, D.; Bonanni, L.; Labrune, J.-B. Radical atoms: Beyond tangible bits, toward transformable materials. *interactions* **2012**, *19*, 38-51.
38. Ishii, H. In *Tangible bits: Beyond pixels*, Proceedings of the 2nd international conference on Tangible and embedded interaction, 2008; ACM: pp xv-xxv.
39. Patten, J.; Ishii, H. In *Mechanical constraints as computational constraints in tabletop tangible interfaces*, Proceedings of the SIGCHI conference on Human factors in computing systems, 2007; ACM: pp 809-818.
40. Lee, J.; Post, R.; Ishii, H. In *Zeron: Mid-air tangible interaction enabled by computer controlled magnetic levitation*, Proceedings of the 24th annual ACM symposium on User interface software and technology, 2011; ACM: pp 327-336.
41. Follmer, S.; Leithinger, D.; Olwal, A.; Hogge, A.; Ishii, H. In *Inform: Dynamic physical affordances and constraints through shape and object actuation*, UIST, 2013; pp 417-426.
42. Raffle, H.S.; Parkes, A.J.; Ishii, H. In *Topobo: A constructive assembly system with kinetic memory*, Proceedings of the SIGCHI conference on Human factors in computing systems, 2004; ACM: pp 647-654.
43. Rasmussen, M.K.; Merritt, T.; Alonso, M.B.; Petersen, M.G. In *Balancing user and system control in shape-changing interfaces: A designerly exploration*, Proceedings of the TEI'16: Tenth International Conference on Tangible, Embedded, and Embodied Interaction, 2016; ACM: pp 202-210.
44. Willett, W.; Jansen, Y.; Dragicevic, P. Embedded data representations. *IEEE Transactions on Visualization and Computer Graphics* **2017**, *23*, 461-470.
45. Jansen, Y.; Dragicevic, P.; Isenberg, P.; Alexander, J.; Karnik, A.; Kildal, J.; Subramanian, S.; Hornb, K.; #230. Opportunities and challenges for data physicalization. In *Proceedings of the 33rd Annual ACM Conference on Human Factors in Computing Systems*, ACM: Seoul, Republic of Korea, 2015; pp 3227-3236.
46. Kranz, M.; Holleis, P.; Schmidt, A. Embedded interaction: Interacting with the internet of things. *Internet Computing, IEEE* **2010**, *14*, 46-53.
47. Guinard, D.; Trifa, V. In *Towards the web of things: Web mashups for embedded devices*, Workshop on Mashups, Enterprise Mashups and Lightweight Composition on the Web (MEM 2009), in proceedings of WWW (International World Wide Web Conferences), Madrid, Spain, 2009.

48. Mayer, S.; Tschofen, A.; Dey, A.K.; Mattern, F. User interfaces for smart things--a generative approach with semantic interaction descriptions. *ACM Transactions on Computer-Human Interaction (TOCHI)* **2014**, *21*, 12.
49. Gallacher, S. Internet of tangible things - talk during the second european tangible interaction studio. http://www.etis.estia.fr/index.php/scientific-aim.html#talk-gallacher (April 2016),
50. European tangible interaction studio (etis). http://www.etis.estia.fr/
51. Ambe, A.H.; Brereton, M.; Soro, A.; Roe, P. In *Technology individuation: The foibles of augmented everyday objects*, Proceedings of the 2017 CHI Conference on Human Factors in Computing Systems, 2017; ACM: pp 6632-6644.
52. Gallacher, S.; Golsteijn, C.; Wall, L.; Koeman, L.; Andberg, S.; Capra, L.; Rogers, Y. In *Getting quizzical about physical: Observing experiences with a tangible questionnaire*, Proceedings of the 2015 ACM International Joint Conference on Pervasive and Ubiquitous Computing, 2015; ACM: pp 263-273.
53. Vaisutis, K.; Brereton, M.; Robertson, T.; Vetere, F.; Durick, J.; Nansen, B.; Buys, L. In *Invisible connections: Investigating older people's emotions and social relations around objects*, Proceedings of the SIGCHI Conference on Human Factors in Computing Systems, 2014; ACM: pp 1937-1940.
54. Grosse-Puppendahl, T.; Herber, S.; Wimmer, R.; Englert, F.; Beck, S.; von Wilmsdorff, J.; Wichert, R.; Kuijper, A. In *Capacitive near-field communication for ubiquitous interaction and perception*, Proceedings of the 2014 ACM International Joint Conference on Pervasive and Ubiquitous Computing, 2014; ACM: pp 231-242.
55. Knutsen, J.; Martinussen, E.S.; Arnall, T.; Morrison, A. Investigating an "internet of hybrid products": Assembling products, interactions, services, and networks through design. *Computers and Composition* **2011**, *28*, 195-204.
56. Van der Vlist, B.J.; Niezen, G.; Hu, J.; Feijs, L.M. In *Semantic connections: Exploring and manipulating connections in smart spaces*, Computers and Communications (ISCC), 2010 IEEE Symposium on, 2010; IEEE: pp 1-4.
57. Domaszewicz, J.; Lalis, S.; Paczesny, T.; Pruszkowski, A.; Ala-Louko, M. Graspable and resource-flexible applications for pervasive computing at home. *IEEE Communications Magazine* **2013**, *51*, 160-169.
58. Lebrun, Y.; Adama, E.; Mandiau, R.; Kolski, C. Interaction between tangible and virtual agents on interactive tables: Principles and case study. *Procedia Computer Science* **2013**, *19*, 32-39.
59. Lebrun, Y.; Adam, E.; Mandiau, R.; Kolski, C. A model for managing interactions between tangible and virtual agents on an rfid interactive tabletop: Case study in traffic simulation. *Journal of Computer and System Sciences* **2015**, *81*, 585-598.
60. Liberati, A.; Altman, D.G.; Tetzlaff, J.; Mulrow, C.; Gøtzsche, P.C.; Ioannidis, J.P.; Clarke, M.; Devereaux, P.J.; Kleijnen, J.; Moher, D. The prisma statement for reporting systematic reviews and meta-analyses of studies that evaluate health care interventions: Explanation and elaboration. *PLoS medicine* **2009**, *6*, e1000100.
61. Scott-Harden, S. In *Active forms for responsive environments*, Proceedings of the 2012 ACM international conference on Intelligent User Interfaces, 2012; ACM: pp 353-358.
62. Matviienko, A.; Ananthanarayan, S.; Heuten, W.; Boll, S. In *Awarekit: Exploring a tangible interaction paradigm for digital calendars*, Proceedings of the 2017 CHI Conference Extended Abstracts on Human Factors in Computing Systems, 2017; ACM: pp 1877-1884.
63. Möller, A.; Roalter, L.; Kranz, M. In *Cognitive objects for human-computer interaction and human-robot interaction*, Human-Robot Interaction (HRI), 2011 6th ACM/IEEE International Conference on, 2011; IEEE: pp 207-208.

64. de la Guía, E.; López, V.; Olivares, T.; Lozano, M.D.; Penichet, V.; Orozco, L. In *Easy smart-home environment to assist patients with mobility impairment*, Proceedings of the 3rd 2015 Workshop on ICTs for improving Patients Rehabilitation Research Techniques, 2015; ACM: pp 122-125.
65. Baraldi, S.; Benini, L.; Cafini, O.; Del Bimbo, A.; Farella, E.; Gelmini, G.; Landucci, L.; Pieracci, A.; Torpei, N. In *Evolving tuis with smart objects for multi-context interaction*, CHI'08 extended abstracts on Human factors in computing systems, 2008; ACM: pp 2955-2960.
66. Ng, R.S.; Kandala, R.; Marie-Foley, S.; Lo, D.; Steenson, M.W.; Lee, A.S. In *Expressing intent: An exploration of rich interactions*, Proceedings of the TEI'16: Tenth International Conference on Tangible, Embedded, and Embodied Interaction, 2016; ACM: pp 524-531.
67. Hannula, P.; Harjuniemi, E.; Napari, E. In *Iot owl: Soft tangible user interface for detecting the presence of people*, Proceedings of the 18th International Conference on Human-Computer Interaction with Mobile Devices and Services Adjunct, 2016; ACM: pp 1168-1172.
68. Krishnaswamy, M.; Lee, B.; Murthy, C.; Rosenfeld, H.; Lee, A.S. In *Iyagi: An immersive storytelling tool for healthy bedtime routine*, Proceedings of the Tenth International Conference on Tangible, Embedded, and Embodied Interaction, 2017; ACM: pp 603-608.
69. Lebrun, Y.; Lepreux, S.; Haudegond, S.; Kolski, C.; Mandiau, R. Management of distributed rfid surfaces: A cooking assistant for ambient computing in kitchen. *Procedia Computer Science* **2014**, *32*, 21-28.
70. Molyneaux, D.; Gellersen, H. In *Projected interfaces: Enabling serendipitous interaction with smart tangible objects*, Proceedings of the 3rd International Conference on Tangible and Embedded Interaction, 2009; ACM: pp 385-392.
71. Mora, S.; Gianni, F.; Divitini, M. In *Rapiot toolkit: Rapid prototyping of collaborative internet of things applications*, Collaboration Technologies and Systems (CTS), 2016 International Conference on, 2016; IEEE: pp 438-445.
72. Bellucci, A.; Vianello, A.; Florack, Y.; Jacucci, G. Supporting the serendipitous use of domestic technologies. *IEEE Pervasive Computing* **2016**, *15*, 16-25.
73. Mora, S.; Divitini, M.; Gianni, F. In *Tiles: An inventor toolkit for interactive objects*, Proceedings of the International Working Conference on Advanced Visual Interfaces, 2016; ACM: pp 332-333.
74. Jacob, R.J.; Girouard, A.; Hirshfield, L.M.; Horn, M.S.; Shaer, O.; Solovey, E.T.; Zigelbaum, J. In *Reality-based interaction: A framework for post-wimp interfaces*, Proceedings of the SIGCHI conference on Human factors in computing systems, 2008; ACM: pp 201-210.
75. Bakker, S.; Antle, A.N.; Van Den Hoven, E. Embodied metaphors in tangible interaction design. *Personal and Ubiquitous Computing* **2012**, *16*, 433-449.
76. Norman, D.A. Natural user interfaces are not natural. *interactions* **2010**, *17*, 6-10.
77. Angelini, L.; Carrino, F.; Carrino, S.; Caon, M.; Khaled, O.A.; Baumgartner, J.; Sonderegger, A.; Lalanne, D.; Mugellini, E. In *Gesturing on the steering wheel: A user-elicited taxonomy*, Proceedings of the 6th International Conference on Automotive User Interfaces and Interactive Vehicular Applications, 2014; ACM: pp 1-8.
78. Valdes, C.; Eastman, D.; Grote, C.; Thatte, S.; Shaer, O.; Mazalek, A.; Ullmer, B.; Konkel, M.K. In *Exploring the design space of gestural interaction with active tokens through user-defined gestures*, Proceedings of the SIGCHI Conference on Human Factors in Computing Systems, 2014; ACM: pp 4107-4116.
79. Angelini, L.; Couture, N.; Khaled, O.A.; Mugellini, E. In *Interaction with the internet of tangible things (iott)*, CEUR Workshop Proceedings of the Third European Tangible Interaction Studio (ETIS), 2017.